\documentclass[twocolumn,prb]{revtex4}
\usepackage{amssymb}
\usepackage{graphicx}
\usepackage{dcolumn}
\usepackage{bm}

\begin{document}
\title{The `Cheerios effect'}
\author{Dominic Vella}
\author{L. Mahadevan}
\email{lm@deas.harvard.edu}
\affiliation{Division of Engineering and Applied Sciences, Harvard University, Pierce Hall, 29 Oxford Street, Cambridge MA 02138}
\begin{abstract}
Objects that float at the interface between a liquid and a gas interact because of interfacial deformation and the effect of gravity. We highlight the crucial role of buoyancy in this interaction, which, for small particles, prevails over the capillary suction that is often assumed to be the dominant effect. We emphasize this point using a simple classroom demonstration, and then derive the physical conditions leading to mutual attraction or repulsion. We also quantify the force of interaction in some particular instances and present a simple dynamical model of this interaction. The results obtained from this model are then validated by comparison to experimental results for the mutual attraction of two identical spherical particles. We conclude by looking at some of the applications of the effect that can be found in the natural and manmade worlds.
\end{abstract}

\date{\today}
\maketitle

\section{\label{intro} Introduction}

\begin{figure}
\centering
\includegraphics[height=4cm]{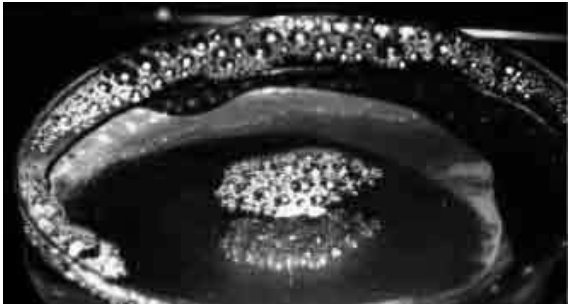}
\caption{Bubbles floating on water in a petri dish. The bubbles are observed to aggregate before moving to the wall of the container. After sufficiently long times, the island of bubbles in the centre also migrates to the wall.}
\label{bubble}
\end{figure}

Bubbles trapped at the interface between a liquid and a gas rarely rest. Over a timescale of several seconds to minutes, long-lived bubbles  move towards one another and, when contained, tend to drift towards the exterior walls - as shown in fig.\ \ref{bubble}. The sceptical reader may readily verify these claims by pouring themselves a glass of sparkling water (or, if they prefer, wine) and following the motion of those bubbles at the surface - particularly those at the periphery of the glass. This phenomenon has even been affectionately dubbed the `Cheerios effect' after the observation that breakfast cereals floating in milk often clump together or stick to the walls of the breakfast bowl \cite{Walker}.

In this article, we bring together a number of perspectives on the Cheerios effect gathered from the literature and our own experience at home, in the kitchen, and in the laboratory. We show how simple physical ideas lead to an understanding not only of the attraction itself, but also of its dynamical consequences. Despite being a subject with enormous potential for simple, reliable party tricks, the technological implications of the Cheerios effect are far from frivolous. Indeed, much research is currently being undertaken to investigate the possibility of using surface tension to induce the `self-assembly' of small-scale structures \cite{Whitesides}. Understanding the way in which particles aggregate at an interface and thence being able to control the form of the aggregate as well as the dynamics of its formation may one day enable much simplified manufacture of components of micro-electromechanical systems (MEMS).

For floating objects in equilibrium or motion, we must consider the balance of linear momentum both in the flotation plane as well as out of the plane, and in addition the balance of angular momentum  in all three directions. Many of the misconceptions in the field arise from considering only some but not all of these balance equations. In particular, neglecting the condition of vertical force balance leads to an underestimation of the importance of the particle's buoyancy in determining the nature of the interaction. 

We begin with a discussion of the physical mechanism that leads to the observed attraction in most instances and illustrate the role of particle buoyancy by means of a simple experiment. We then focus on a series of simple examples that allow us to quantify the magnitude of the attractive force. The first of these is inspired by an oversimplified physical picture that is often perceived as the dominant effect. By considering carefully the vertical force balance that must be satisfied for particles to float, we then show that it is \textit{this} that is dominant for small particles and subsequently propose a simple dynamical model for the attraction of two spherical particles. Finally, we show that consideration of the remaining equilibrium condition, that of torque balance, can lead to amphiphilic strips and conclude with some possible biological implications of our observations.

\section{\label{physics} The physical origin of attraction}

\begin{figure}
\centering
\includegraphics[height=3.5cm]{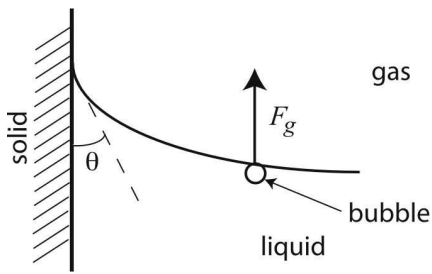}
\caption{Schematic of a single bubble close to a wall, along with the definition of the contact angle.}
\label{defn}
\end{figure}

The mechanism behind the apparent attraction between bubbles or between a bubble and the wall of a glass is easy to understand by considering the geometry of the interface at which the bubbles are trapped. For simplicity, we consider the latter case (schematically illustrated in fig.\ \ref{defn}), although the explanation of the clustering of many bubbles is similar. Here, the air-water interface is significantly distorted by the presence of the wall (the well-known meniscus effect) and, since the bubble is positively buoyant, there is a net upward force, $F_g$, on the bubble. Because it is constrained to lie at the interface, however, the bubble cannot simply rise vertically and instead does the next best thing by moving upwards \emph{along} the meniscus. As glass is in general wetting (the contact angle $\theta$, defined in fig.\ \ref{defn}, satisfies $\theta< \pi/2$), in moving upwards along the meniscus the bubble also moves closer to the wall. Viewed from above, it appears as if there is an exotic attractive force acting between the wall and the bubble when in fact the bubble is merely obeying gravity and moving in a constrained path imposed by the presence of the wall.

Now, a single bubble will deform the interface just as the presence of a wall does, though this time for a different reason and to a lesser extent. In the bubble's case, it can only remain at the interface because the buoyancy force (that is tending to push the bubble out of the liquid) is counterbalanced by the surface tension force (that opposes the deformation of the interface and hence is tending to keep the bubble in the liquid). These two competing effects reach a compromise where the bubble is partially out of the liquid but the interface is slightly deformed. This deformation is significant enough to influence other bubbles nearby, which move `upwards' along the meniscus, and so spontaneously aggregate.

This mechanism was first proposed by Nicolson \cite{Nicolson} as a means by which the bubbles that constitute a bubble raft interact and give the raft its solid-like properties. As we shall see in section \ref{chan}, this mechanism provides the dominant contribution for the interaction between sufficiently small particles.

\begin{figure}
\centering
\includegraphics[height=4.5cm]{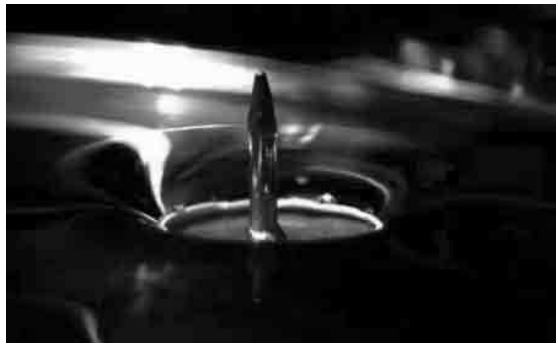}
\caption{A photograph of a drawing pin floating upturned on water. Notice that the deformation of the interface in this case is the opposite to that around a bubble or near a wetting wall.}
\label{pin}
\end{figure}

The same argument works in the case of particles that are significantly more dense than water floating at an interface. That such heavy particles can float at all is due to the fact that this time surface tension stops the interface from deforming too much \emph{downwards} as would happen if the particle were to sink. This reversal of the interfacial curvature can be seen clearly in fig.\ \ref{pin} for a metal drawing pin floating on water: surface tension must act upwards to counterbalance the weight of the pin. By analogy with what is observed with bubbles, we should expect that another drawing pin floating near the first will `fall' down the interface and hence the two appear to be attracted to one another - as is indeed observed.

\section{repulsion: often misunderstood}

So far we have seen only that the deformation of an interface caused by the presence of particles at that interface can lead to mutual attraction between those particles and eventually to the formation of large clusters. This, however, is only half of the story. Imagine that we were to float a buoyant bubble in the vicinity of a drawing pin - should they also attract? On the basis of the previous physical argument, we expect that the bubble will move upwards along the interface distorted by the presence of the drawing pin. However, since the pin is not buoyant (i.e. the interface has the curvature shown in fig.\ \ref{pin}), moving along the interface will, in this case, cause the bubble to move away from the drawing pin and so the two objects in fact \emph{repel} one another.

 A more striking demonstration of this repulsion can be achieved using two drawing pins, provided they are the kind that has a thin plastic cap around the blunt end. As we should expect from the discussion given in section \ref{physics}, these two drawing pins will attract when floated at the interface. However, if we now carefully remove the plastic cap from the top of one and float it (the cap) near the intact drawing pin, then the two will be seen to move apart.

This simple experiment apparently challenges the common assumption that the attraction or repulsion of particles at interfaces should depend solely on the wetting properties, namely the contact angles, of the particles (see, for example, p. 70, ex. 3 of Batchelor's book \cite{Batchelor} or Campbell \textit{et al.} \cite{Campbell}). Here, the wetting properties of the plastic cap (which is the only part that is in contact with the liquid) are not altered by removing it from the pin but the weight that it must support is much reduced, making the cap buoyant. In turn, this alters the balance between surface tension and gravity so that the interface must now pull down on the cap to keep it at the interface and so the deformation near the cap resembles that around a bubble. This change in the sign of the curvature of the interface has been brought about without changing the surface properties of the cap, simply because of a change in the effective density of the particle: a possibility with many potential applications that appear not to have been explored fully, although it is well known to those who know well \cite{Kralchevsky}.

\section{\label{platecalc} A model calculation}

\begin{figure}
\centering
\includegraphics[height=6cm]{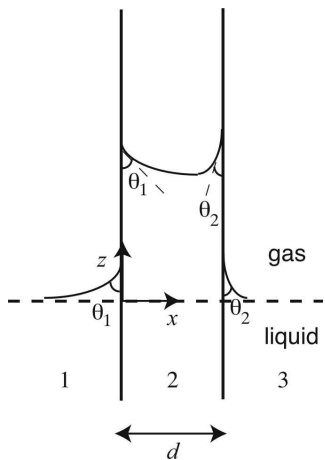}
\caption{The geometry of two infinite plates in a semi-infinite
fluid. The planes have contact angles $\theta_1$ and $\theta_2$
respectively and are a horizontal distance $d$ apart.}
\label{plates}
\end{figure}

Quantifying the physical picture outlined in section \ref{physics} allows us not only to predict the conditions under which the interfacial curvature changes sign, but also to understand at a simple level the  dynamical interaction between two particles.  We start with an idealised problem in which we account only for the condition of horizontal force balance (and neglect the vertical force and torque balance conditions) by focusing on two infinite vertical plates at a liquid-gas interface, as shown in fig.\ \ref{plates}. The presence of the plates distorts the interface leading to a force of attraction between the two plates whose magnitude we shall now calculate. This set-up has, in fact, been used as a model for explaining the Cheerios effect \cite{Walker} and while we argue later that this picture is incorrect for floating objects, it does lend itself to a simple calculation (for a further simplification leading to similar conclusions, the reader is referred to the Encyclopedia Brittanica article by Clerk-Maxwell \cite{Maxwell}).

The equation of the interface $z=h(x)$ is determined from the condition that the pressure change across the interface due to surface tension (which is proportional to the curvature of the interface) is equal to the hydrostatic pressure difference caused by the deformation of the interface (see p. 65 of Batchelor \cite{Batchelor} for a thorough discussion of this). For small interfacial deflections, this balance may be written as:
\begin{equation}
\gamma \frac{d ^2h}{d x^2}=\rho gh
\label{plateeqn}
\end{equation} where $\gamma$ is the surface tension coefficient of the liquid-gas interface, $\rho$ is the density of the liquid and $g$ is the acceleration due to gravity. Equation (\ref{plateeqn}) is to be solved with the boundary conditions that the contact angles $\theta_1$ and $\theta_2$ are given at each of the plates and that the deflection of the interface should decay far away from the plates. In each of the regions labelled $i=1,2, 3$ in fig.\ \ref{plates}, the solution of (\ref{plateeqn}) gives the equation of the interface to be $z=h_i(x)=A_ie^{-x/L_c}+B_ie^{x/L_c}$ where $L_c\equiv\sqrt{\gamma/\rho g}$ is the \emph{capillary length}, which defines the length-scale over which interactions occur. The conditions $h_1(-\infty)=0=h_3(\infty)$ give $A_1=0=B_3$, which, combined with the contact angle conditions $h_1'(0)=\cot\theta_1$ and $h_3'(d)=-\cot\theta_2$, give the interface shape outside the plates as:
\begin{eqnarray}
\frac{h_1(x)}{L_c}=\cot\theta_1 e^{x/L_c}\\
\frac{h_3(x)}{L_c}=\cot\theta_2 e^{(d-x)/L_c}
\end{eqnarray} while the contact angle conditions $h_2'(0)=-\cot\theta_1$ and $h_2'(d)=\cot\theta_2$ give the interface shape between the two plates to be:
\begin{equation}
\frac{h_2(x)}{L_c}=\frac{\cot\theta_1\cosh\left( \frac{d-x}{L_c}\right)+\cot\theta_2\cosh(x/L_c)}{\sinh(d/L_c)}
\end{equation} 

\begin{figure}
\centering
\includegraphics[height=4.5cm]{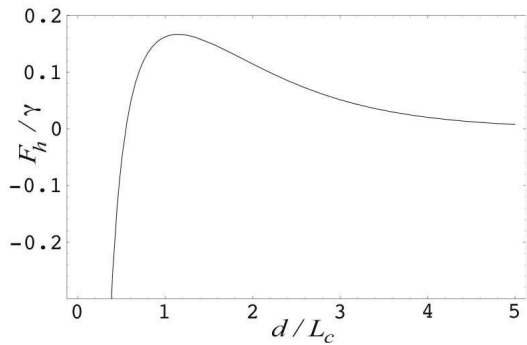}
\caption{A typical force-separation curve for a wetting and non-wetting plate and a liquid-gas interface. Here, $\theta_1=2\pi/3$ and $\theta_2=\pi/4$ and we observe repulsion at large separations as well as attraction at short range.}
\label{fplatecurv}
\end{figure}

Because of the interfacial deformation given by (2)-(4), the plates are now subjected to a capillary pressure that results in a horizontal force on the plates (there is no resultant horizontal surface tension force because its components on either side of a plate cancel exactly). This force may act either to bring them together or to pull them apart depending on the two contact angles $\theta_1$ and $\theta_2$. The value of the horizontal force per unit length, $F_h$, can be calculated by integrating the hydrostatic pressure along each of the wetted sides of one of the plates (say the one on the left in fig.\ \ref{plates}) and taking the difference as follows:
\begin{eqnarray}
F_h&=&-\int_{-\infty}^{h_2(0)}\rho g z dz+\int_{-\infty}^{h_1(0)}\rho g z dz=\int^{h_1(0)}_{h_2(0)}\rho g z dz\nonumber \\
&=&\frac{1}{2} \rho g(h_1(0)^2-h_2(0)^2)
\label{capforce}
\end{eqnarray} so that we have:
\begin{equation}
F_h=-\frac{\gamma}{2} \left(\frac{(\cot\theta_1\cosh(d/L_c)+\cot\theta_2)^2}{\sinh^2(d/L_c)}-\cot^2\theta_1\right)
\label{fplate}
\end{equation} where the sign convention is such that $F_h<0$ corresponds to attraction between the plates. Typically this force is either attractive for all plate separations or repulsive at large separations and attractive at short separations (with an unstable equilibrium at an intermediate distance). An example of a force-displacement curve in the latter case is shown in fig.\ \ref{fplatecurv}. Equation (\ref{fplate}) can be used to show that repulsion is possible only if $\cot\theta_1\cot\theta_2<0$, i.e. if one plate is wetting and the other non-wetting, as follows. First, we note that, provided that $\cot\theta_1\neq-\cot\theta_2$, the fact that $F_h\rightarrow 0$ as $d\rightarrow\infty$ implies that repulsion can only occur if $F_h$ has a maximum value somewhere, since as $d\rightarrow 0$, $F_h\rightarrow-\infty$. A simple calculation shows that the only turning point of the function $f(\xi)=(\cot\theta_1\cosh\xi+\cot\theta_2)/\sinh\xi$ occurs at $\xi=\xi^*$ where $\cosh\xi^*=-\cot\theta_2/\cot\theta_1$, which only has a real solution $\xi^*$ provided that $\cot\theta_1\cot\theta_2<0$. When $\cot\theta_1=-\cot\theta_2$, the short range attraction does not exist and instead there is repulsion at \emph{all} displacements, but the result that repulsion can only occur when $\cot\theta_1\cot\theta_2<0$ still stands.

\begin{figure}
\centering
\includegraphics[height=6cm]{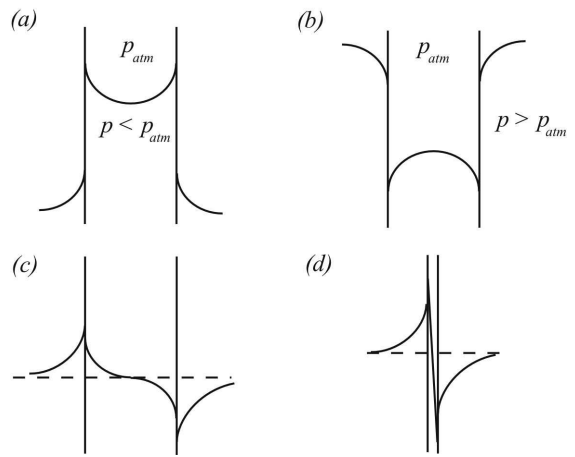}
\caption{Typical interfacial profiles for the different configurations of plate wettabilities. (a) Two wetting plates. (b) Two non-wetting plates. (c) A wetting and non-wetting plate at intermediate separation. (d) A wetting and non-wetting plate at short range.}
\label{expl}
\end{figure}

The result just derived shows that vertical plates at a liquid-gas interface will attract if they have like menisci and repel otherwise, as we saw in section \ref{physics} with floating objects. However, the physical mechanism here is subtly different. We can no longer argue in terms of one plate following the meniscus imposed by the other as these plates do not float, meaning that there is no analogue of the condition of vertical force balance in this scenario. Instead, we must consider the effects of hydrostatic pressure that result from the deformation of the interface, as detailed by Poynting and Thomson \cite{Poynting}
. Here, we give an abbreviated version of their argument in terms of the configurations shown in fig.\ \ref{expl} (a) and (b) in which plates of like wettability are at the interface. In (a), the upper portion of the central column of fluid is at a pressure lower than atmospheric pressure, $p_{atm}$, because of the curvature of the interface and the two plates attract because of the excess of atmospheric pressure. Similarly in (b), the pressure in the outer fluid is greater than $p_{atm}$, again because of the sign of the interfacial curvature, and so there is an excess pressure (or capillary suction) causing the two plates to attract. The situation is more complicated when one plate is wetting and the other non-wetting. At intermediate and large displacements, the interfacial displacement of the central column at the points where it touches a plate is \emph{smaller} than it is on the other side of the same plate (as shown in fig.\ \ref{expl} (c)) due to the constraint that the central interface must pass through $z=0$ (rather than just being asymptotic to $0$) to allow it to satisfy both contact angle conditions. From (\ref{capforce}), this is enough to show that the two plates repel. When the two plates come close to contact, however, a relatively large change in the gradient of the intermediate meniscus is required between the two plates, inducing a large curvature. Since the curvature of the interface is proportional to its height, this in turn means that the displacement of the interface must be large in this region, reversing the sign of the force and leading to attraction (see fig.\ \ref{expl} (d)). The exception to this is the case where $\cot\theta_1=-\cot\theta_2$, as noted in the discussion following (\ref{fplate}). Here, very little curvature is required since the contact angles are precisely complementary and so, in turn, very little displacement of the interface is necessary. Thus, the interfacial displacement of the inside of the plates is smaller than that on the outside of the plates and so, again using (\ref{capforce}), we see mutual repulsion between the two plates, regardless of the displacement between them.

The argument just given has often been invoked to explain the Cheerios effect (see, for example, the answer to problem 3.100 of Walker's book \cite{Walker} or perform a Google search on `Cheerios effect'). While the effects of hydrostatic pressure imbalances caused by interfacial deformation are certainly important, the calculation considered in this section cannot constitute a complete explanation since it neglects the crucial fact that Cheerios, and other floating objects, must satisfy a vertical force balance to be able to float. As we shall see in the next section, this changes the physics fundamentally for small particles, leading to the buoyancy mechanism outlined in section \ref{physics} providing the dominant effect. Incorporating this additional equilibrium condition, however, requires a slightly more involved calculation, which we shall now consider.

\section{\label{chan} The case of floating objects}

Having seen the manner in which the force between two interfacial objects can be calculated in a somewhat artificial geometry, we are now in a position to consider the scenario that is of most interest to us here: interactions between objects that are \emph{floating} at a liquid-gas interface. The major difference between this and what has gone before is that we must now take into account the vertical force equilibrium of the object. This is actually a significant enough complication that, even using the linearised approach of section \ref{platecalc}, progress can only be made numerically. However, for sufficiently small particles we may assume in addition that the interfacial profiles generated by one or more object floating at the interface are sufficiently small that they may be superposed and so make progress analytically.

This assumption was tacitly made in our discussion of the attraction and repulsion of objects at interfaces in section \ref{physics}, but was first introduced by Nicolson \cite{Nicolson} who used it to calculate the interaction force between neighbouring bubbles in bubble rafts. It was then applied to the calculation of the force between floating particles at an interface by Chan \textit{et al.} \cite{Chan} who considered some simple illustrative particle configurations such as two horizontal cylinders floating near one another. Here we give their argument applied to the problem of determining the interaction between two identical spheres floating at an interface, primarily motivated by the simplicity of an experimental realization thereof.

\begin{figure}
\centering
\includegraphics[height=3.5cm]{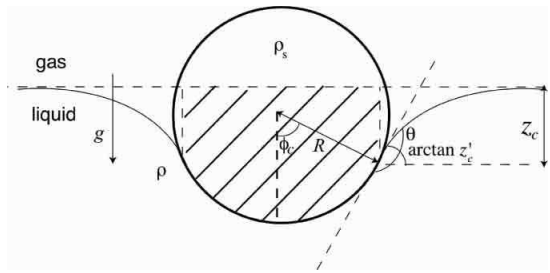}
\caption{Geometry of a sphere lying at a liquid-gas interface. The shaded area represents the weight of liquid equivalent to the buoyancy force due to hydrostatic pressure acting on the sphere \cite{Keller}.}
\label{chanfig}
\end{figure}

\subsection{A single particle}

In the spirit of the Nicolson approximation, we first neglect the presence of the second sphere and consider the vertical force balance on an isolated sphere to determine an approximate expression for the interfacial slope at the contact point, $z_c'$ (see fig.\ \ref{chanfig}). This is equivalent to determining the value of $\phi_c$, as defined in fig.\ \ref{chanfig}, because of the geometrical relation $\phi_c=\pi-\theta+\arctan z_c'$, but the calculation of $ z_c'$ will prove useful to justify Nicolson's approximation and to compare the results with those of section \ref{platecalc}. 

For this sphere to remain at the interface, its weight, $\frac{4}{3}\pi\rho_sg R^3$ must be balanced by the component of surface tension acting along the (circular) contact line and the buoyancy force because of the displaced bulk fluid. The first of these is easily seen to be given by $2\pi R\sin\phi_c\gamma\sin(\arctan z_c')=2\pi \gamma R\sin\phi_c z_c'(1+ z_c'^2)^{-1/2}$, while the second is just given by the weight of water that \emph{would} occupy the area between the wetted region of the sphere and the undisturbed interface, which is shown as the hatched area in fig.\ \ref{chanfig}. To understand physically this generalisation of Archimedes principle, notice that the liquid has no knowledge of the geometry of the object that is at the interface outside of its wetted perimeter. The liquid must therefore produce an upthrust equal to that it would provide to an object filling the entire hatched region, which we know from the usual Archimedes result is just the weight of displaced liquid that would fill this volume. (For elegant rigourous derivations of this result, the reader might consult Keller \cite{Keller} or Mansfield \textit{et al.} \cite{Mansfield}). This volume can be calculated by splitting it into a circular cylinder of radius $R\sin\phi_c$ and height $z_c$ and a spherical cap of height $R(1-\cos\phi_c)$ and base $R\sin\phi_c$. This gives the buoyancy force to be: 
\begin{equation}
\pi\rho gR^3\left(\frac{z_c}{R}\sin^2\phi_c+\frac{2}{3}-\cos\phi_c+\frac{1}{3}\cos^3\phi_c\right)
\end{equation}The balance of these vertical forces may now be written explicitly as:
\begin{eqnarray}
\label{vforcebal}
\frac{4}{3}\pi\rho_sg R^3&=&2\pi\gamma R\sin\phi_c\frac{ z_c'}{\sqrt{1+ z_c'^2}}\\
&+&\rho g\pi R^3\left( \frac{z_c}{R}\sin^2\phi_c+\frac{2}{3} -\cos\phi_c+\frac{1}{3}\cos^3\phi_c\right) \nonumber
\end{eqnarray} By substituting $\phi_c=\pi-\theta+\arctan{ z_c'}$ and keeping only those terms linear in $ z_c'$ we obtain an expression for $ z_c'\sin\phi_c$ accurate to linear order in the \emph{Bond number}, $B\equiv R^2/L_c^2$:
\begin{equation}
 z_c'\sin\phi_c=B\left(\frac{2D-1}{3}-\frac{1}{2}\cos\theta+\frac{1}{6}\cos^3\theta\right)\equiv B\Sigma
\label{changrad}
\end{equation} where $D\equiv \rho_s/\rho$. (As a consistency check, observe that $ z_c'=0$ when $\theta=\pi/2$ and $D=1/2$, which we expect, since in this case the Archimedes buoyancy alone is enough to balance the weight of the sphere without any interfacial deformation.)

Equation (\ref{changrad}) contains two dimensionless parameters, $B$ and $\Sigma$, whose physical significance it is important to discuss. The Bond number, $B=\rho gR^2/\gamma$, is the most important dimensionless parameter in this system. Physically, it gives a measure of the relative importance of the effects of gravity and surface tension: large $B$ corresponds to large particles or small surface tension coefficient - in both cases surface tension is inconsequential. The expression for the slope of the interface in the vicinity of the spherical particle given in (\ref{changrad}) is valid for $B\ll1$ (corresponding to a radius of $\sim1$mm or smaller for a sphere at an air-water interface) in which case surface tension is very important. The other dimensionless parameter, $\Sigma$, can be thought of as a (non-dimensional) resultant weight of the particle once the Archimedes upthrust has been subtracted out. This physical interpretation arises naturally from the vertical force balance condition (\ref{vforcebal}) and (\ref{changrad}) since the resultant weight of the object (in the linearised approximation) is simply $2\pi \gamma R z_c'\sin\phi_c=2\pi \gamma RB\Sigma$.

To calculate the interaction energy using the Nicolson approximation, we must also calculate the interfacial displacement caused by an isolated floating sphere, which is determined by the hydrostatic balance $\gamma \nabla^2h=\rho gh$ - the co-ordinate invariant statement of equation (\ref{plateeqn}). With the assumption of cylindrical symmetry, this becomes:
\begin{equation}
\frac{1}{r}\frac{d}{dr}\left(r\frac{dh}{dr}\right)=\frac{h}{L_c^2}
\label{besseleqn}
\end{equation} with the boundary conditions that $h\rightarrow0$ as $r\rightarrow\infty$ and $h'(r=R\sin\phi_c)= z_c'$. Equation (\ref{besseleqn}) has a solution in terms of modified Bessel functions of the first kind\cite{Abramowitz} and of order $i$, $K_i(x)$:
\begin{equation}
h(r)=- z_c'L_c\frac{K_0(r/L_c)}{K_1(R\sin\phi_c/L_c)}\approx - z_c'\sin\phi_c RK_0(r/L_c)
\label{intprof}
\end{equation} where we have used the asymptotic result (see, for example, Abramowitz and Stegun\cite{Abramowitz}) that $K_1(x)\approx 1/x$ for $x\ll1$ to simplify the prefactor. 

\subsection{Two interacting particles}

Having calculated the effective weight of a sphere at a deformed interface as $2\pi \gamma RB\Sigma$ (with $\Sigma$ as defined in equation (\ref{changrad})) as well as the interfacial deformation caused by the presence of a single sphere we are now in a position to calculate the energy of interaction between two spheres. To leading order in $B$, this energy is the product of the resultant weight of one sphere and its vertical displacement caused by the presence of another sphere with its centre a horizontal distance $l$ away. We may therefore write the energy, $E(l)$, as:
\begin{equation}
E(l)=-2\pi\gamma R^2B^2\Sigma^2K_0\left(\frac{l}{L_c}\right)
\label{sphenergy}
\end{equation} and from this, the force of interaction is given by $F(l)=-\frac{dE}{dl}$ or:
\begin{equation}
F(l)=-2\pi\gamma RB^{5/2}\Sigma^2K_1\left(\frac{l}{L_c}\right)
\label{sphereforce}
\end{equation}

\begin{center}\underline{\hspace{7cm}}\end{center}
\paragraph*{\textbf{Exercise:}} Carry out the above calculation for the case of two cylinders of infinite length lying horizontally and parallel to one another at an interface. To do this, first consider the interfacial profile caused by an isolated cylinder and show that this is given by 
\begin{equation}
z(x)=-L_cz_c'\exp(-x/L_c)
\label{cylprof}
\end{equation} when $B\ll1$. Next use the linearised vertical force balance and the geometrical relationship $\phi_c=\pi-\theta+\arctan z_c'$ to show that:
\begin{equation}
z_c'\approx\frac{B}{2}\left( \pi(D-1)+\theta-\frac{1}{2}\sin2\theta \right)\equiv\frac{BC}{2}
\label{cyllaw}
\end{equation} The resultant weight of the object can be found from the force balance to be $\approx2\gamma z_c'$. Use this and the interfacial profile in (\ref{cylprof}) to calculate the energy of interaction (per unit length) between two cylinders with centre-centre separation $l$, $E(l)$. Show that:
\begin{equation} 
F(l)=-\frac{dE}{dl}=-\frac{\gamma}{2}B^2C^2\exp(-\frac{l}{L_c})
\end{equation} with $C$ defined as in (\ref{cyllaw}).
\begin{center}\underline{\hspace{7cm}}\end{center}

It is important to emphasise that the calculation leading to (\ref{sphenergy}) relied on a number of assumptions. The first of these is that the particle be small enough that the total interfacial deformation is the sum of that due to individual particles (by comparison with numerical results, Chan \textit{et al.} \cite{Chan} show that the expression derived by their method is essentially exact for Bond numbers $B\leq 0.1$). Furthermore, we have neglected the effect of capillary pressure acting on the particle to produce a horizontal force - calculated for the case of two vertical plates in section \ref{platecalc}. It is not possible using the analysis presented here to include this effect since inherent in the Nicolson approximation is the assumption that the level of the interface is the same on either side of the particle. However, the difference in interface heights on either side of the sphere only occurs at the next order in $B$ so that using (\ref{fplate}), we see that the contribution from capillary pressure also only enters at the next order in $B$ without doing the detailed calculation. 

At large Bond numbers the analysis presented above breaks down raising the question: is it still the case that large particles interact because of their gravitational potential energy or does the capillary suction effect discussed in section \ref{platecalc} become more important? To answer this question conclusively requires the numerical solution of the full problem since for large Bond number the interfacial deflections are no longer small; a calculation that is beyond the scope of this work. Such a calculation has, however, been performed by Allain and Cloitre \cite{Allain}, who showed that for $B\gg 1$ the vertical displacement of two horizontal cylinders does not change substantially as they move towards one another. This in turn means that the attractive force must result largely from pressure effects rather than the weight of the particles, with the crossover between these two regimes occurring for $10<B < 100$ according to Allain and Cloitre's numerical results. The toroidal shape of a Cheerio complicates the notion of Bond number for a Cheerio but if we take an effective radius based on its volume $R^*=(R_1^2R_2)^{1/3}\approx 2.7$mm (where $R_1\approx2$mm and $R_2\approx5$mm are the two radii of the torus) and $L_c=2.7$mm for an air-water interface then $B\approx1$. This is within the regime where the gravitational energy of the particles dominates the capillary suction due to the meniscus between them and so it is crucial that we account for the buoyancy effects to correctly interpret the attractive force.

Finally, we discuss briefly how the result in (\ref{sphereforce}) fits together with the physical interpretation of attraction and repulsion that we developed in section \ref{physics}. If they perform the exercise below, the reader will see that the sign of the interaction force between two non-identical spheres is governed by the sign of $ z_c'^{(1)}\sin\phi_c^{(1)}\times z_c'^{(2)}\sin\phi_c^{(2)}$, where the superscript $(i)$ labels the two particles. In particular, the interaction is attractive if this product is positive and repulsive if it is negative. From (\ref{intprof}), we see that this product has the same sign as the product of the gradients of the menisci in the neighbourhood of the two particles, and so we see that there is mutual attraction if the particles cause like menisci and repulsion if they have unlike menisci. This is precisely the picture that we saw in section \ref{physics}, although we are now able to quantify the combination of contact angles and particle densities that give rise to the two different possibilities. We also note that the strength of the interaction \emph{decreases} as the surface tension coefficient, $\gamma$, increases. This slightly counterintuitive result is actually a simple consequence of the fact that for higher values of $\gamma$ the deformation of the interface required to satisfy the vertical force balance condition for particle 1 is less and so the gravitational hill upon which particle 2 finds itself is smaller.

\begin{center}\underline{\hspace{7cm}}\end{center}
\paragraph*{\textbf{Exercise:}} By repeating the analysis for two spheres but now allowing them to have different material properties $R$, $\theta$ and $D$, show that the sign of the interaction force for two (non-identical) spheres is determined by the sign of $ z_c'^{(1)}\sin\phi_c^{(1)}\times z_c'^{(2)}\sin\phi_c^{(2)}$, where the superscript $(i)$ labels the two particles (the interaction is attractive if this product is positive and repulsive if it is negative).
\begin{center}\underline{\hspace{7cm}}\end{center}

\section{\label{dynamics} the dynamics of floating particles }

Having so far limited ourselves to calculating the force of interaction between particles, we shall now use this  to answer some questions that arise naturally from observing the motion of objects at an interface. Perhaps the most natural question to start with is ``how fast do two spherical particles come together?" and it is this that we shall focus on in this section. 

To make the dynamical problem tractable, we assume that the vertical velocity of each particle as it moves along the meniscus is small enough that the vertical force balance used to determine the horizontal force is satisfied; an assumption that is valid to leading order, as shown by Chan \textit{et al.} \cite{Chan}. 

\begin{figure}
\centering
\includegraphics[height=4.5cm]{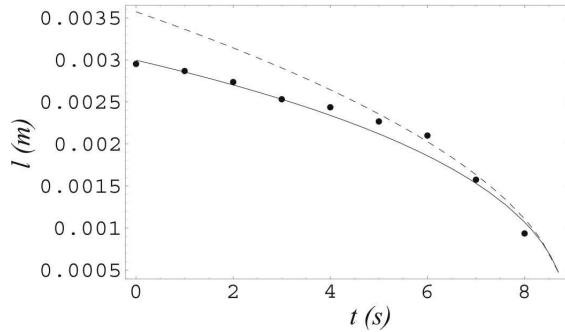}
\caption{ Experimental data (points) compared to the solutions of the dynamical equation (17) (solid line) for two identical spheres of radius $0.3$mm interacting via flotation forces at an air-water interface. The theoretical curve is calculated by solving  (17) numerically with $\gamma=0.0728$ Nm$^{-1}$, $\mu=0.001$ Pa s and $\Sigma^2/\alpha=0.673$ used as the fitting parameter. ($\Sigma$ is the non-dimensional resultant weight of the spheres and $\alpha$ is the ratio of the drag that the particle feels at the interface to that which it would experience in an unbounded fluid). The dashed line gives the asymptotic result (\ref{asysol}).}
\label{spheredat}
\end{figure}

We assume that the motion is overdamped so that a Stokes drag term \cite{Batchelor} (for the viscous drag provided by the liquid) balances the attractive force between the two particles given in (\ref{sphereforce}). This balance leads to the following equation of motion:
\begin{equation}
6\pi\mu R \alpha \frac{dl}{dt}=-2\pi\gamma R B^{5/2}\Sigma^2K_1\left(\frac{l(t)}{L_c}\right)
\label{overdamped}
\end{equation} where $\mu$ is the dynamic viscosity of the liquid and $\alpha$ is a scaling factor to take into account the fact that the drag experienced by a particle at an interface is less than it would experience if completely immersed in the bulk fluid. We expect that $\alpha\approx1/2$ although Danov \textit{et al.} \cite{Danov} computed numerically the dependence of $\alpha$ on $\theta$. Data obtained from observing two spherical particles with radius $0.3$ mm as they move under each other's influence at the interface between air and water is shown in fig.\ \ref{spheredat}. This data was collected from a time lapse movie (one frame per second) of the motion taken with a digital camcorder, which was then analysed using image analysis software (ImageJ, NIH). Here the non-dimensional resultant weight of the particles, $\Sigma$, is difficult to measure experimentally  because of its dependence on the contact angle $\theta$. We thus appear to have two unknown parameters in this model ($\alpha$ and $\Sigma$). Fortunately, since only the ratio $\Sigma^2/\alpha$ appears in (\ref{overdamped}) we can fit the numerical solution of (\ref{overdamped}) to the experimental data presented in fig.\ \ref{spheredat} via this one parameter.

In this case, we are assuming that the particles are sufficiently small that their inertia may be neglected entirely. In fact, the approach we have adopted is better suited to such small particles since the Bond number in this situation is also very small and so the expression for the interaction force in (\ref{sphereforce}) is effectively exact. Also, if the typical distance between particles is small (compared to the capillary length, $L_c$) then we are again able to use the asymptotic formula $K_1(x)\approx x^{-1}$ for $x\ll1$ to approximate the modified Bessel function that occurs in the force-law (\ref{sphereforce}). Substituting this into (\ref{overdamped}) allows us to solve for $l(t)$, giving:
\begin{equation}
l(t)\approx\sqrt{l(0)^2-\frac{2\gamma L_cB^{5/2}\Sigma^2}{3\mu\alpha}t}
\label{asysol}
\end{equation} so that the time taken for two spheres to come into contact is given by:
\begin{equation}
t_{contact}\approx\frac{3\mu\alpha(l(0)^2-R^2)}{2\gamma L_cB^{5/2}\Sigma^2}
\end{equation}
In fig. 8, we see that the asymptotic form (\ref{asysol}) is a reasonable approximation over the last 2 seconds before contact.

To verify our assumption that the motion is slow enough for us to be able to neglect the particle's acceleration, we look at the ratio of particle inertia to Stokes drag. This ratio ${\cal R}$ is initially small and increases as the particles come closer, reaching a maximum when they come in contact. Using the asymptotic expression for $K_1(x)$, we have:
\begin{equation}
{\cal R}_{contact}=\frac{2}{27\alpha^2}\frac{\gamma\rho L_cB^{5/2}\Sigma^2}{3\mu^2}
\end{equation} For spherical particles of radius less than $0.3$mm (or $B=0.01$), ${\cal R}<0.1$ throughout the motion, which is sufficiently small that our approach is self-consistent. 

\section{Amphiphilic Strips \textit{without} chemistry}

\begin{figure}
\centering
\vspace{0.5cm}
\includegraphics[height=4cm]{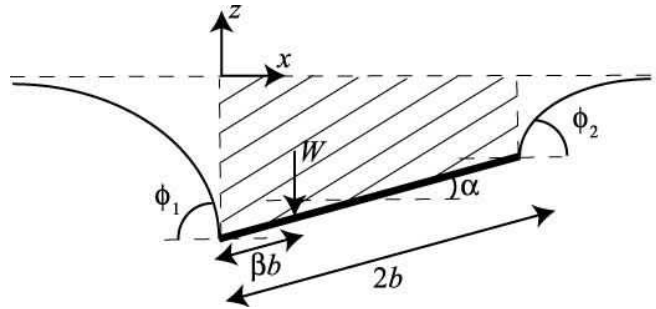}
\caption{A single two-dimensional strip (of weight $W$ per unit length in the direction perpendicular to the page) floating at an interface, with the hatched area indicating the area of displaced fluid, the weight of which is equal to the buoyancy force on the strip. The asymmetric equilibrium position here is possible because of an off-centre position for the strip's centre of mass.}
\label{amph1}
\end{figure}

As a final illustration of the principles that we have applied to the problem of the attraction of interfacial objects we shall consider briefly the equilibrium of a single strip (of weight $W$ per unit length) floating horizontally at an interface, as shown in fig.\ \ref{amph1}. This problem was studied extensively by Mansfield \textit{et al.} \cite{Mansfield} who also investigated the interactions between two such strips. Here, however, we content ourselves with studying the simpler problem of a single strip but with a slight twist: we take the centre of mass line of the strip to be displaced from the strip's centre line by a distance $(1-\beta)b$. As we shall see, this is enough to break the symmetry of the problem and thus allow the strip to float at an angle $\alpha$ to the horizontal.

Because the strip is assumed to be infinitely thin, the concept of contact angle that we met earlier is not well-defined in this problem. Instead, the effective contact angles that the interface makes at the points of contact with the strip, $\phi_{1,2}$, and the angle $\alpha$ are determined from the condition that the strip be in equilibrium, which leads to three equations for the three unknowns. The first two of these are the conditions of horizontal and vertical force balance, which we have encountered previously and here may be written:
\begin{equation}
0=\gamma(\cos\phi_1-\cos\phi_2)+\frac{\rho g}{2}(z_1^2-z_2^2)
\label{horeq}
\end{equation} for horizontal equilibrium and:
\begin{equation}
W=\gamma(\sin\phi_1+\sin\phi_2)-\rho g(2b\cos\alpha)\left(\frac{z_1+z_2}{2}\right)
\label{vereq}
\end{equation} for vertical equilibrium - the second term on the right hand side in (\ref{vereq}) simply being the weight of liquid displaced by the strip, which is shown as the hatched area in fig.\ \ref{amph1}. In addition, there is the condition of torque balance, which we did not encounter for the equilibrium of the floating sphere (it is automatically satisfied for objects of circular cross-section). Here, however, this condition is crucial as it gives a third equation by which to determine the three unknowns. By taking moments about the centre of mass, we have:
\begin{eqnarray}
0&=&\gamma b\left( (2-\beta)\sin(\phi_2-\alpha)-\beta\sin(\phi_1+\alpha)\right) \nonumber\\ 
&-&\rho g\int_{-\beta b}^{(2-\beta)b} z(s)s ds
\label{torbal}
\end{eqnarray} where $s$ is an arc-length co-ordinate measured along the strip from the centre of mass. 

In principle, these equations can be solved even for interfacial deformations that are not small following the strategy outlined by Mansfield \textit{et al.} \cite{Mansfield}. However, for simplicity, we proceed here in the limit where the three angles are small and, therefore, the interfacial deformations are also small. In this limit $z_{1}\approx -L_c\phi_{1}, z_{2}\approx -L_c\phi_{2}$  so that the vertical relation (\ref{vereq}) becomes:
\begin{equation}
\frac{W}{\gamma}=(\phi_1+\phi_2)\left(1+\frac{b}{L_c}\right)
\label{linveq}
\end{equation} Then, the linearized version of (\ref{horeq}) is automatically satisfied and so we make use of the geometrical relation $z_2-z_1=2b\sin\alpha$ to give:
\begin{equation}
\alpha=\frac{\phi_1-\phi_2}{2b/L_c}
\end{equation} Finally, the torque balance condition (\ref{torbal}) yields:
\begin{eqnarray}
0&=&(2-\beta)\phi_2-\beta\phi_1-2\alpha\nonumber\\
&+&\frac{b}{L_c}(1-\beta)(\beta\phi_2+(2-\beta)\phi_1)\nonumber\\
&+&\frac{b}{L_c}\frac{\phi_2-\phi_1}{6}(\beta^3+(2-\beta)^3)
\label{lintor}
\end{eqnarray} 

\begin{figure}
\centering
\vspace{0.5cm}
\includegraphics[height=4.5cm]{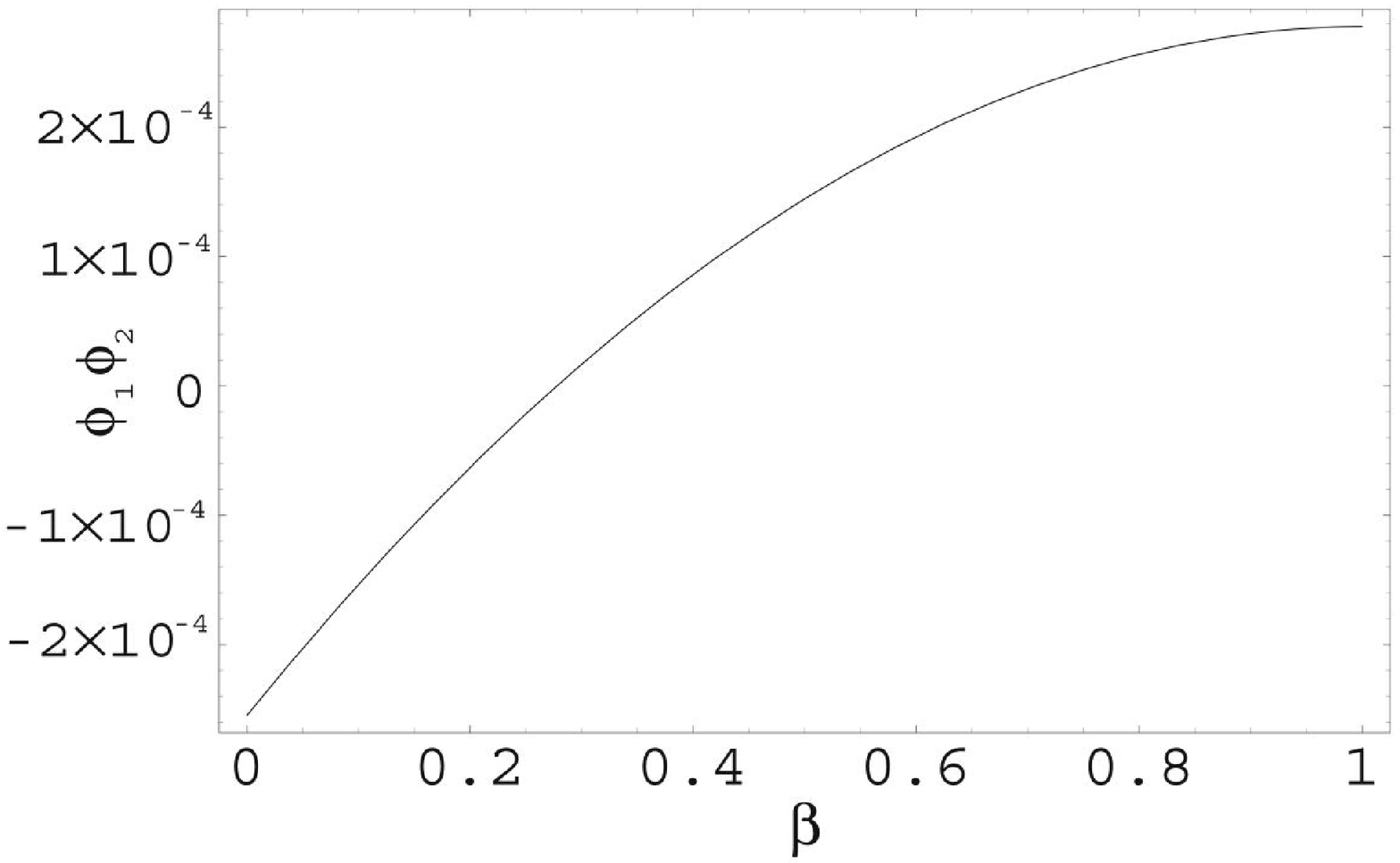}
\caption{The dependence of $\phi_1\phi_2$ on the offset of the strip's centre of mass, given by $(1-\beta)b$, for $\beta\in [0,1]$ (the graph being symmetric about $\beta=1$). Here the strip has non-dimensional weight per unit length $W/\gamma=0.1$ and half-width $b/L_c=2$.}
\label{amphi}
\end{figure}

Together, equations (\ref{linveq})-(\ref{lintor}) constitute a system of three linear equations in three unknowns, which can be solved simply by inverting a $3\times3$ matrix . Of particular interest here is the possibility that, for some values of $\beta$, we may have $\phi_1$ and $\phi_2$ of opposite sign and so plotted in fig.\ \ref{amphi} is the dependence of $\phi_1\phi_2$ on $\beta$ showing that for sufficiently small $\beta$, $\phi_1$ and $\phi_2$ are of opposite sign. We can understand this more fully by solving the system (\ref{linveq})-(\ref{lintor}) to give:
\begin{widetext}
\begin{equation}
\phi_1\phi_2=\frac{W^2}{\gamma^2}L_c^2\frac{(3L_c^2+6bL_c+4b^2-3b(L_c+b)\beta)(3L_c^2-2b^2+3b(L_c+b)\beta)}{4(L_c+b)^2(b^2+3bL_c+3L_c^2)^2}
\end{equation}
\end{widetext} From this and the restriction that $\beta\leq1$, it is a simple matter to show that $\phi_1\phi_2<0$ when $\beta<\beta_c=\frac{2b^2-3L_c^2}{3b(b+L_c)}$ and hence that $\phi_1$ and $\phi_2$ are of opposite sign for sufficiently small $\beta$ provided that $b>(3/2)^{1/2}L_c$. For the parameters in fig. \ref{amphi}, $\beta_c\approx0.28$, with $\phi_2$ changing from negative to positive around $\beta=\beta_c$ and $\phi_1$ positive for all $\beta\in[0,1]$. Physically, this is what we should expect since for small $\beta$, the offset of the centre of mass causes a large torque and hence rotation of the object, forcing the far end of the strip to be displaced above the equilibrium liquid level and hence the interface here must be elevated, corresponding to $\phi_2<0$. As the offset decreases, so does the torque associated with it and the displacement of the far end of the strip is diminished until, for sufficiently small offsets, the far end lies below the equilibrium liquid level and $\phi_2>0$.

The significance of the qualitatively different interfacial shapes at either end of the strip that we see when $\phi_1\phi_2<0$ is that in such cases one edge is able to attract drawing pins while the other repels drawing pins and attracts bubbles! Particles or molecules with this type of behaviour are often termed \emph{amphiphiles} and occur in detergents along with many other applications \cite{Ondarcuhu}. However, such amphiphilic particles are usually constructed by treating the two edges chemically to induce the very different behaviour: here, we have shown that this may also be achieved by altering the density of the strip so that the centre of mass of the object is displaced.

\begin{figure}
\centering
\vspace{0.5cm}
\includegraphics[height=4.5cm]{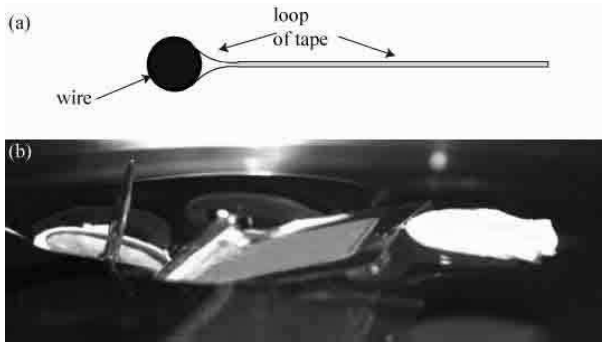}
\caption{Example of a simple amphiphilic strip. (a) Simple design of the amphiphilic strip mentioned in the text. (b) Photograph of such an amphihilic strip bonding a drawing pin (left) and the cap of another drawing pin (right), which would normally be mutually repulsive. Using such an amphiphile allows particles that would otherwise experience mutual repulsion to be maintained stably at an almost arbitrary separation.}
\label{amphipic}
\end{figure}

As well as the possible industrial applications that particles such as this could have, this \textit{physical} amphiphile lends itself to a much simpler classroom demonstration than might be possible with chemical amphiphiles. Using a piece of sticky tape doubled back on itself to form the strip, see fig.\ \ref{amphipic} (a),  and a short length of wire (part of a paperclip for example) inserted between the two sides of tape, it is possible to make a strip with edges that deform the interface in manifestly different ways. An experimental realization of this is shown in fig.\ \ref{amphipic} (b), which demonstrates that such particles are an effective means by which particles that would otherwise be mutually repulsive can be coaxed into maintaining a finite equilibrium separation.

\section{Discussion}

In this article we have investigated an aspect of everyday life that may previously have escaped our readers' notice - the propensity of floating objects to aggregate. Having couched the mechanism for this effect in terms of the simple physics of particles trapped at a deformed interface feeling the effects of their weight (or buoyancy), we have shown how approximate methods can lead to quantitative descriptions of the magnitude and dynamical nature of the interaction. We now conclude with a brief discussion of some of the scenarios in which this effect has found application and point out some potential research directions.

\begin{figure}
\centering
\vspace{0.5cm}
\includegraphics[height=5cm]{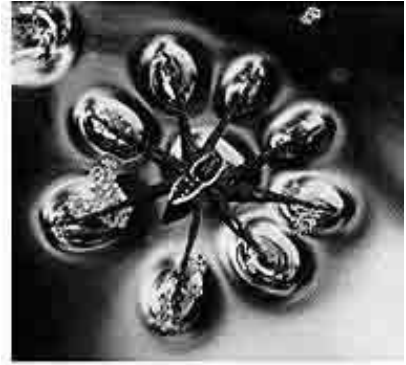}
\caption{The interfacial deformation caused by a water-spider (\textit{Dolomedes triton}). Here, the interface is visibly depressed by the spider's weight acting on each leg. Image courtesy of Robert B. Suter, Vassar College.}
\label{spider}
\end{figure}

As is often the case, the first place to look beyond the kitchen and the lab is the natural world. There are likely many instances where a variant of the Cheerios effect is used by some species or another. Here, however, we choose to highlight one particular application: water walking creatures that can be found on the surface of many ponds. Many of these animals rely on surface tension to prevent them from drowning as their weight can be supported by interfacial deformations (as is shown in fig.\ \ref{spider}). However, when they try to climb out of the pond, they become reluctant victims of the Cheerios effect  since this generally requires climbing up the meniscus - against gravity. Recent observations  by Hu and  Bush \cite{Hu} suggest that some insects, such as \textit{Mniovelia Kuscheli},  may overcome this difficulty by pulling up on the interface with their front legs and pushing down on it with their hind legs, effectively shifting their center of gravity. They thus become the natural exemplar of the mechanical amphiphile discussed in the last section as they are now attracted to the wall via the Cheerios effect.

There remain many interesting questions that we have not been able to answer in this article, many of which are certainly amenable to investigation in the classroom or the laboratory. For example, it would also be worthwhile to better understand the way in which two different types of particles that are mutually repulsive interact. With a mixture of light and heavy particles, for example, clusters of like density particles form - all things (other than the particle density) being equal. This segregation phenomenon would undoubtedly find many applications within science and industry. Such a study could then naturally be extended by the inclusion of amphiphilic particles, which would allow the user to dictate a finite equilibrium spacing between two otherwise repulsive objects. Another example is afforded by hair on water: a flexible hair floating parallel to a planar wall might be expected to bend as it is attracted to the wall since the attractive force would be greatest for those parts of the hair closest to the wall and much less significant for those further away. This motion has yet to be studied from either an experimental or theoretical point of view.  The inspiration from the kitchen, industry and nature seems almost overwhelming. 

\acknowledgments

This article arose as a result of a summer studentship funded by the Heilbronn Fund of Trinity College, Cambridge (D.V.). We thank John Bush and David Hu for telling us about the meniscus-climbing insect \textit{Mniovelia Kuscheli}, Robert Suter for permission to use his photograph in fig. \ref{spider} and  Pascale Aussillous for assistance with the photography in figures \ref{bubble}, \ref{pin} and \ref{amphipic} (b). The anonymous reviewers also made many useful suggestions.

\end{document}